\documentclass[aps,prd,showpacs,twocolumn]{revtex4}

\usepackage{epsfig}
\usepackage{longtable}
\usepackage{subfigure}

\begin{document}

\title{Unravelling an Extra Neutral Gauge Boson at the LHC Using Third Generation Fermions}
\author{ Ross Diener$^{1,2}$, Stephen Godfrey$^{1,3}$\footnote{Email: godfrey@physics.carleton.ca} 
and Travis A. W. Martin$^1$\footnote{Email: tmartin@physics.carleton.ca}}
\affiliation{
$^1$Ottawa-Carleton Institute for Physics, Department of Physics, Carleton University, Ottawa, Canada 
K1S 5B6\\
$^2$ Perimeter Institute for Theoretical Physics, 31 Caroline St N, Waterloo, Canada N2L 2Y5 \\
$^3$TRIUMF, 4004 Wesbrook Mall, Vancouver, Canada V6T 2A3 }

\date{\today}

\begin{abstract}

We study the potential to use measurements of the properties of extra neutral gauge bosons ($Z^{\prime}$'s) 
in $pp$ collisions at the Large Hadron Collider to unravel the underlying physics.  
We focus on the usefulness of third generation final states 
($\tau$, $b$, $t$) in distinguishing between models with non-universal $Z^{\prime}$-fermion couplings.
We present an update of discovery limits of $Z^{\prime}$'s including the 2010-2011 LHC run 
and include models with non-universal couplings.
We show how ratios of $\sigma(pp \to Z^{\prime} \to t\bar{t})$, 
$\sigma(pp \to Z^{\prime} \to b\bar{b})$, and $\sigma(pp \to Z^{\prime} \to \tau^+\tau^-)$ 
to $\sigma(pp\to Z^{\prime} \to \mu^+\mu^-)$ can be used to distinguish between models and 
measure parameters of the models.  Of specific interest are models with preferential couplings
such as models with generation dependent couplings.
We also find that forward-backward asymmetry measurements with third generation fermions in the final state could provide important input to understanding the nature of the $Z^\prime$. Understanding detector resolution and efficiencies will be crucial for extracting results.
\end{abstract}

\maketitle

\section{Introduction}

With the startup of the Large Hadron Collider (LHC), it is possible that direct 
experimental evidence of new physics will soon follow.
Many theoretical models of physics beyond the Standard Model
include a new colourless, spin-1, neutral gauge boson 
($Z^{\prime}$)~\cite{Langacker:2008yv, Leike:1998wr, Cvetic:1995zs, Hewett:1988xc, Rizzo:2006nw}.
A kinematically accessible $Z^{\prime}$ would be one of the most distinctive signals at the LHC, 
and may potentially be one of the earliest discoveries.
If a heavy neutral gauge boson were discovered, the immediate task would be to understand its 
origins by measuring its properties.
This subject has been explored extensively 
and continues to be an important topic of on-going research 
\cite{Petriello:2008zr,Godfrey:2008vf,Diener:2009ee,Diener:2009vq,Li:2009xh,delAguila:2009gz,
Barger:2009xg,Osland:2009tn,Rizzo:2009pu}.

The ATLAS~\cite{ATLAS:2008zzm} and CMS~\cite{Ball:2007zza} 
experiments at the LHC expect to be able to identify $\tau$ leptons and $b$- and $t$-quark 
jets in $pp$ collisions. 
In a recent letter, we suggested that third generation fermions can provide an important tool 
for discriminating between candidate models of $Z^{\prime}$'s~\cite{Godfrey:2008vf}.
Along this vein, we found that such measurements can also be used to explore the gauge 
symmetry of some models of new physics that include a $Z^{\prime}$ with distinctive properties
reflecting the mixing between the underlying gauge groups. 
These include models with fermion type dependent couplings (i.e., preference for quarks
over leptons)~\cite{Georgi:1989xz,Chivukula:1994qw, Chivukula:2002ry}, 
or generation dependent couplings (i.e., preference for third generation) 
\cite{Bardeen:1989ds, Chivukula:1994mn, Chivukula:1995gu, Simmons:1996ws, Malkawi:1996fs, 
Muller:1996qs, He:1999vp, Hill:1994hp, Lane:1995gw, Lane:1996ua, Lane:1998qi, Popovic:1998vb}.
The models with generation dependent couplings, such as extended technicolour
\cite{Chivukula:1994mn, Chivukula:1995gu, Simmons:1996ws} 
and topcolor assisted technicolour 
\cite{Hill:1994hp, Lane:1995gw, Lane:1996ua, Lane:1998qi, Popovic:1998vb}, 
are especially interesting as they 
take the heavy top quark mass as evidence for new physics that is connected to the 
mechanism of electroweak symmetry breaking (EWSB).  Thus, evidence for the violation of generation 
universality could give important hints to the mechanism of EWSB.
For such models, standard analysis methods that rely on Drell-Yan processes 
with either electron or muon final states may fail to identify important details
of the underlying physics.

In this paper, we look at a number of measurements involving third generation final states, 
expanding on our previous letter~\cite{Godfrey:2008vf}.
We start by exploring the discovery reach for the models we consider.
However, our main focus is to consider the viability and usefulness of 
measurements involving third generation fermions in identifying the underlying 
theory of a newly discovered $Z^{\prime}$.
The measurements we study are the ratio of the third generation quark cross sections 
to the muon cross section, the ratio of the tau cross section to the muon cross section, 
and the forward-backward asymmetry with bottom and top-quark final states.

We found that the ratios of the branching fraction of $\tau^+ \tau^-$, $b\bar{b}$ and $t\bar{t}$ 
final states to the 
$\mu^+ \mu^-$ final state in $Z^{\prime}$ decays are especially useful for understanding models with 
preferential couplings and can be used to quantify the extent of the preference.
We also found that $A_{FB}$ measurements with third generation quarks in the final state have 
the potential to provide $Z^{\prime}$-fermion coupling information.
Several papers have been published in the past  on related topics~\cite{Rizzo:1993yx, Han:2005ru, 
Lynch:2000md, Harris:1999ya, Baur:2008uv, Frederix:2007gi, Mohapatra:1992tc, Agashe:2007ki,
Chekanov:2010vc}.
This paper takes into account recent information on the expected capability of the 
ATLAS~\cite{ATLAS:2008zzm} and CMS~\cite{Ball:2007zza}  detectors in addition to backgrounds.

We begin in Section II with a brief survey of models with non-universal couplings 
to $Z^{\prime}$'s, followed in Section III with details of our calculations, focusing 
on issues of third generation fermion identification and Standard Model backgrounds.
In Section IV we give our results, starting with an update of discovery limits for 
$Z^{\prime}$'s from the models we consider in this paper and reflecting the updated 
LHC running plans.
Our main focus, however, is exploring how one can use third generation fermions 
to learn about the underlying theory that gives rise to a $Z^{\prime}$.
Finally,  we summarize our main conclusions in Section V.

\section{Models}

New massive resonances are present in many models, including KK theories with finite size 
extra dimensions, string theories and theories with extended gauge sectors.
In this paper we focus on models with extended gauge sectors.
Some examples of these models are variations of Little Higgs Models (Littlest Higgs (LH)
with $\tan\phi_{H} = 1.0$~\cite{ArkaniHamed:2002qy}, 
Simplest Little Higgs (SLH)~\cite{Schmaltz:2004de}, 
Anomaly Free Simple Little Higgs (AFSLH)~\cite{Kaplan:2003uc}),  
$E_{6}$ models ($\chi$, $\psi$, $\eta$), 
Left Right Symmetric models (LRM, ALRM) ($g_{R} = g_{L}$), and 3-3-1 models~\cite{Pisano:1991ee}.
The details of the $E6$ and LR Symmetric models have been described elsewhere, so
we refer the interested reader to the literature
\cite{Langacker:2008yv, Leike:1998wr, Cvetic:1995zs, Hewett:1988xc, Rizzo:2006nw}.

We single out a subset of models where spontaneous breaking to diagonal subgroups 
results in generators corresponding to gauge bosons that couple to either different 
generations (first vs third) 
or different types (quark vs lepton) of fermions with differing strengths.
These models typically have a group structure of the form $SU(2) \times SU(2)$ 
or $U(1) \times U(1)$.
Such models include the Ununified Standard Model ~\cite{Georgi:1989xz}, 
non-commuting extended technicolor~\cite{Chivukula:1994mn}, 
top-flavour~\cite{Malkawi:1996fs, Muller:1996qs, He:1999vp}, 
and topcolor assisted technicolor models~\cite{Hill:1994hp}.
These models can be distinguished using third generation fermions and we therefore 
give some relevant details of the models that are used in our analysis.

\subsection{Ununified Model (UUM)}

In the Ununified Model~\cite{Georgi:1989xz}
the left handed quarks and leptons transform as doublets under their 
respective $SU(2)$ groups in $SU(2)_{q} \times SU(2)_{l} \times U(1)_{Y}$. 
Right handed fermions transform as singlets under both groups, and hypercharge assignments 
remain the same as in the SM.

After symmetry breaking, the mass eigenstates for the gauge bosons include 
a massless photon, and two massive $Z$ bosons.
The light $Z^0$ boson deviates from SM couplings at order $\sin^{2}\phi$, where $\phi$ is 
the mixing parameter between the two $SU(2)$ groups and is expected to be small.
The heavy boson then couples to the lepton and quark sectors as:
\begin{equation}
g_{Z^{\prime}} = g_{Z^{0}} c_{w} \left( \frac{T_{3q}}{\tan\phi_{UUM}} - \tan\phi_{UUM} T_{3l} \right).
\end{equation}
Chivukula \textit{et al.}~\cite{Chivukula:1994qw, Chivukula:2002ry} 
found 95\% C.L. constraints on the value of 
$M_{Z^{\prime}}$ dependent on the mixing angle.
In general, $M_{Z^{\prime}}>2$~TeV is required, 
with a stronger mass constraint for larger values of $\sin^{2}\phi_{UUM}$.
In our study,
the use of $M_{Z^{\prime}} = 1.5$~TeV is for comparison to other models without such constraints.
Unless otherwise stated, we take $\sin\phi_{UUM} = 0.5$, following the work of 
Chivukula \textit{et al.}~\cite{Chivukula:1994qw, Chivukula:2002ry}.

\subsection{$SU(2)_{h} \times SU(2)_{l}$ - Extended Technicolor (ETC)}

In models of an extended $SU(2)_{h} \times SU(2)_{l}$ structure, fermion generations 
transform differently under each gauge group - the first two generations transform as a doublet 
under the  $SU(2)_{l}$, while the third generation transforms as a doublet under the $SU(2)_{h}$.
Subsequently, this extended gauge group is broken to its familiar diagonal subgroup, $SU(2)_{L}$, 
at some energy scale $\mu$.
The electric charge operator is given by:
\begin{equation}
Q = T_{3l} + T_{3h} + Y.
\end{equation}

In diagonalizing the mass matrix for the neutral gauge bosons, a nearly SM $Z^{0}$ arises with 
a $Z^{\prime}$ that couples only to left handed fermions.
This $Z^{\prime}$ then has a greatly enhanced coupling strength to third generation fermions, 
with 
couplings given by:
\begin{equation}
g_{L} = g_{Z^{0}}c_{w} \left( -\tan\phi_{ETC} T_{3h} + \cot\phi_{ETC} T_{3l} \right).
\end{equation}
Unless otherwise stated, a value of $\sin\phi_{ETC} = 0.9$ is used.
Constraints found by Chivukula \textit{et al.},~\cite{Chivukula:1995gu,Chivukula:2002ry} 
require $M_{Z^{\prime}}>2$~TeV for this value of $\sin\phi_{ETC}$.
Again, calculations are performed with lower values of the $M_{Z^{\prime}}$ for 
comparisons with other models.

Examples of these models include top-flavour models~\cite{Malkawi:1996fs, Muller:1996qs, He:1999vp} and non-commuting extended technicolor 
\cite{Chivukula:1994mn, Chivukula:1995gu, Simmons:1996ws}.

\subsection{$U(1)_{h} \times U(1)_{l}$ - Topcolor Assisted Technicolor (TC2)}

The large mass of the  top quark has led to suggestions that the top quark is intrinsically related 
to the dynamics of electroweak symmetry breaking.
Models based on this idea are often called Topcolor~\cite{Hill:1991at,Hill:2002ap}, 
and contain a
gauge structure that is generation dependent.
It is assumed that the QCD gauge group and the hypercharge $U(1)$ arises from the breaking of a 
larger group as in 
$SU(3)_{1} \times SU(3)_{2} \times U(1)_{1} \times U(1)_{2} \to SU(3)_{c} \times U(1)_{Y}$ 
with a residual, high scale $SU(3)' \times U(1)'$.
In this type of model, the $Z^\prime$ plays a role in the generation of the large top quark mass by 
providing a tilting mechanism for the top quark seesaw~\cite{Harris:1999ya}.

The third generation fermions transform under the $U(1)_1$ group, 
while the first two generation fermions transform under the $U(1)_2$ group.
After symmetry breaking, the resulting $Z^{\prime}$ 
couples differently to the third generation fermions than it does to the first two generation fermions.
The couplings of the SM fermions to the $Z^{\prime}$ are given by:
\begin{eqnarray}
g_{L,R}^{3} = \frac{1}{2}g_{Z^{0}} s_{w} Y_{SM} \cot\phi_{TC2}\\
g_{L,R}^{1,2} = \frac{1}{2}g_{Z^{0}} s_{w} Y_{SM} \tan\phi_{TC2},
\end{eqnarray}
where $Y_{SM}$ is the standard model hypercharge value.
Unless otherwise stated, a value of $\sin\phi_{TC2} = 0.5$ is used.

Examples of these models also include top quark seesaw models~\cite{Hill:1994hp, Lane:1995gw, 
Lane:1996ua} and  flavour-universal TC2~\cite{Lane:1998qi, Popovic:1998vb}.

\section{Calculations}

For our calculations we use the leading order Drell-Yan cross section.
While it can be found in the literature, we include it here for completeness~\cite{Godfrey:1987qz,
Rizzo:2006nw,Dittmar:2003ir}:

\begin{widetext}
\begin{equation}
{{d\sigma}\over {d\cos\hat{\theta}}} \left(pp \to Z^{\prime},Z^{0},\gamma \to f\bar{f} \right) = \sum_q \int 
dx_a dx_b f_q(x_a,Q^2) f_{\bar{q}}(x_b,Q^2) 
{{d\hat{\sigma}(\hat{\theta})}\over{d \cos\hat{\theta}}} 
+ f_{\bar{q}}(x_a,Q^2) f_q(x_b,Q^2) {{d\hat{\sigma}(\pi-\hat{\theta})}\over{d \cos\hat{\theta}}},
\label{eq:zp-cs}
\end{equation}
\\
where $d\sigma/d\cos\theta$ is given by
\begin{equation}
\frac{d\hat{\sigma}}{d\cos\hat{\theta}} \left(q\bar{q} \to Z^{\prime},Z^{0},\gamma \to f\bar{f} \right)= 
\frac{\pi\alpha^{2}_{em}\beta_{f}}{8c^{4}_{W}s^{4}_{W}\hat{s}}
\left\{\left(1+\beta_{f}^{2}\cos^{2}\hat{\theta}\right)S_{q} + 2\beta_{f} 
\cos\hat{\theta}A_{q} + S'_{q}\right\}
\label{eq:diff_cross}
\end{equation}
\\ 
\\
and
\begin{eqnarray}
S_{q},A_{q} = \sum_{i,j=\gamma,Z,Z^{\prime}}\left(\frac{\hat{s}}{\hat{s} - 
M^{2}_{i} - i\Gamma_{i}M_{i}}\right)\left(\frac{\hat{s}}{\hat{s} - M^{2}_{j} 
+ i\Gamma_{j}M_{j}}\right)\times
 \left(R_{f}^{i}R_{f}^{j} \pm L_{f}^{i}L_{f}^{j}\right)\left(R_{q}^{i}R_{q}^{j} \pm L_{q}^{i}L_{q}^{j}\right)\\
S'_{q} = \sum_{i,j=\gamma,Z,Z^{\prime}}\left(\frac{4m^{2}_{f}}{\hat{s}}\right)
\left(\frac{\hat{s}}{\hat{s} - M^{2}_{i} - i\Gamma_{i}M_{i}}\right)
\left(\frac{\hat{s}}{\hat{s} - M^{2}_{j} + i\Gamma_{j}M_{j}}\right)\times 
\left(R_{f}^{i}L_{f}^{j} + L_{f}^{i}R_{f}^{j}\right)
\left(R_{q}^{i}R_{q}^{j} + L_{q}^{i}L_{q}^{j}\right).
\label{eq:sym_asym}
\end{eqnarray}
\end{widetext} 
In Eq.~(\ref{eq:zp-cs}) through (\ref{eq:sym_asym}), 
$M_i$ and $\Gamma_i$ are the masses and widths of the photon, SM $Z^0$, and $Z^{\prime}$; $L^i_f$, 
$R^i_f$ are the left and right handed couplings of the gauge bosons to fermion species $f$;
$m_f$ is the mass of the final state fermion;  $f_{q,\bar{q}}(x,Q^2)$
are the parton distribution functions (pdf's);
$x_a$ and $x_b$ are the momentum fractions of the partons; 
$Q^2$ is the scale at which the parton distribution functions are evaluated, which 
we take to be $\hat{s}$ - the square of the parton centre-of-mass energy;
$\hat{\theta}$ is the centre-of-mass scattering angle; 
and $\beta_f=\sqrt{1-4m_f^2/\hat{s}}$.  
In our calculations we used
$\alpha=1/127.9$, $\sin^2\theta_w=0.231$, $M_Z=91.188$~GeV, $\Gamma_Z=2.495$~GeV and 
$m_t=171.2$~GeV~\cite{Amsler:2008zzb}, and for the pdf's we used set CTEQ6M~\cite{Pumplin:2002vw}.
We calculated the $Z^{\prime}$ width including only $Z^{/prime}$ decays to SM fermions, 
and neglected $Z-Z^{\prime}$ mixing and decays to pairs of $W$ and $Z^0$ bosons,
as well as decays to heavy, fourth generation fermions.

The total cross section 
is proportional to the $S_{q}$ and $S'_{q}$ terms (with summed couplings), 
which are dependent on symmetric combinations of the $Z^{\prime}$-fermion couplings. 
The antisymmetric combinations of the $Z^{\prime}$-fermion couplings in the $A_{q}$ term 
contributes to the forward-backward asymmetry, $A_{FB}$.
Thus,  measuring the production cross section 
and forward-backward asymmetry of the decay of a $Z^{\prime}$ to third generation fermions will 
give complementary information about the $Z^{\prime}$ couplings to fermions - information that 
will be crucial for disentangling the underlying theory. 
The mass dependent coefficient for $S'_{q}$ is only relevant for top decays 
when the mass of the $Z^{\prime}$ is relatively small.
The effect is $O(10\%)$ for a 1~TeV $Z^{\prime}$, dropping to $O(1\%)$ for a 2~TeV $Z^{\prime}$, 
for top quark final states.

In our calculations, we include a K-factor~\cite{KubarAndre:1978uy} 
in the cross section to account for 
NLO QCD corrections and neglect NNLO as the uncertainties in parton distributions 
dominate over such small effects~\cite{Melnikov:2006kv, Anastasiou:2003ds}.
We also included QCD~\cite{Kataev:1992dg, Gorishnii:1988bc, D'Agostini:1989cz} 
and EW~\cite{Baur:2001ze, Baur:1997wa, Baur:2002fn} corrections in the width, 
with weak corrections having only a very minor effect.

The formula given in Eq.~\ref{eq:zp-cs} 
was used to calculate the cross sections and distributions
in this paper.  The phase space integrals were performed using Monte-Carlo integration methods
with weighted events.
QCD backgrounds are calculated using a combination of the WHiZaRD (with O'MEGA matrix elements) 
\cite{Kilian:2007gr, Ohl:2006ae, Kilian:2001qz} and MADGRAPH~\cite{Maltoni:2002qb} 
Monte-Carlo event generators and compared with our own code following Barger and Phillips
\cite{Barger:1987nn}.  All three background calculations agreed.

\section{Identification and Backgrounds}

The challenge to using third generation fermions in LHC events is to first identify the third generation
fermions and then to identify the signal events buried in the large Standard Model QCD backgrounds.
The backgrounds take two forms.
The first is the SM production of third generation quarks which we will discuss below.
The second is the misidentification of the large QCD light jet backgrounds as heavy fermions.
Identification of third generation fermions and misidentification of light jet backgrounds are not unrelated.
The rejection of light jet backgrounds is crucial to making accurate measurements of heavy quark final 
states.
However, higher rejection typically results in lower identification efficiencies, and correspondingly lower 
statistics.
There is therefore a tradeoff between high identification efficiency and suppression of the reducible 
background.

There has been recent activity on this subject with a number of papers appearing on the 
topic of the identification of top jets, specifically with techniques for highly boosted jets
\cite{Kaplan:2008ie,Thaler:2008ju,Baur:2007ck,Baur:2008uv}.
In addition, ATLAS has released updated information on the expected capabilities of the 
detector for the tagging of leptons and bottom quarks~\cite{ATLAS:2008zzm}.
We use this 
information to estimate the efficiency and rejection rates needed to 
analyze the data.
It should be noted that some estimates are given for tagging lower energy events 
and we do not evaluate the validity of extending or extrapolating to higher energies.
Better understanding of the ATLAS and CMS detectors will ultimately give more reliable values.
Finally, there are non-QCD backgrounds such as $W+jets$, $(Wb + W\bar{b})$, and 
$Wb\bar{b} +jets$.
However, studies have shown that these can be controlled by constraints on cluster 
transverse mass and the invariant mass of jets~\cite{Baur:2008uv}
so we will not consider them further.

In the following subsections we summarize the identification efficiencies and fake rejection
rates for the third generation fermions that we use in our analysis.

\subsection{Muon}

Muons are the most distinct of all tagged signatures for both ATLAS and CMS~\cite{Kortner:2007qj}.
Recent studies by the ATLAS Muon Working Group~\cite{ATLAS:2008zzm} show a single muon 
identification efficiency greater than 95\% for $p_{T}>30$~GeV.
These values vary over pseudorapidity and energy; for the purpose of this analysis, we use an 
efficiency of $\epsilon_{\mu} = 96\%$ for single muon identification and consider only Drell-Yan 
backgrounds.

\subsection{Tau Lepton}

There are three possible decay modes for a $\tau^+\tau^-$ event: purely leptonic 
($\tau^+\tau^- \to \nu_\tau \bar{\nu}_\tau l^+\nu l^-\bar{\nu}$, 12.4\% of total events), 
semi-leptonic events ($\tau^+\tau^- \to \nu_\tau \bar{\nu}_\tau l\nu + jets$, 45.6\% of 
total events), and purely hadronic events ($\tau^+\tau^- \to \nu_\tau \bar{\nu}_\tau + jets$, 
42\% of total events). 
Of the hadronic decays, $\approx\!77\%$ have one charged pion track (one-prong) and 
$\approx\!23\%$ have three charged pion tracks (three-prong).
In our analysis, we consider the results from the purely hadronic mode, which suffers 
from a large reducible dijet background, as the leptonic modes have a larger missing energy 
component. It has been shown that a $\tau^+\tau^-$ invariant 
mass distribution can be reconstructed in these modes even though the $\tau$ decays produce 
missing energy~\cite{Rainwater:1998kj,Holdom:2008xx}.  

In order to reduce the dijet background, it will be necessary to implement tight tagging methods.
In particular, ATLAS~\cite{ATLAS:2008zzm} estimates that 
it may be possible to achieve a rejection rate of $10^{3}$ and an efficiency of 20\% for 
3 prong decays with $E_{T}>100$~GeV, and a rejection rate of $10^{3}$ and an 
efficiency of 50\% for 1 prong decays with $E_{T}>100$~GeV.
This amounts to an overall rejection rate of $5 \times 10^{5}$ with $\approx\!20\%$ of all hadronic (42\%)
$\tau^{+}\tau^{-}$ events passing the selection criteria. 
The number of useful $\tau^+\tau^-$ pairs can potentially be increased by including the leptonic
and semileptonic modes.

\subsection{Bottom}

According to the current ATLAS algorithms~\cite{ATLAS:2008zzm}, a jet is tagged as a $b$-jet if a $b$-
quark is found with $p_{T}>5$~GeV within $\Delta R = 0.3$ around the centroid of the jet.
Further improvements are made to this tag by including either the identification of a 
secondary vertex or else reconstructing the invariant mass of the jet.
In an analysis of $b$-jets arising from $WH$ and $t\bar{t}$, a jet rejection of about 
$2 \times 10^{2}$ was found for a $b$-jet efficiency of 60\% - or 36\% for a $b\bar{b}$ pair, 
with a light jet rejection of $4 \times 10^4$.
Tighter tagging could be employed to improve the rejection rate if the light jet background dominates the 
observed events.

\subsection{Top}

A complete analysis of physics events involving top quarks would require an analysis of the decay 
products to calculate an overall efficiency.
This is beyond the scope of this paper and we instead use previous studies to estimate an overall
efficiency~\cite{Kaplan:2008ie, Baur:2008uv, Barger:2006hm, Wagner:2005jh, Almeida:2010pa}.

We considered both the semi-leptonic decay mode
($t\bar{t} \to (l\nu_{l})(jj)b\bar{b}$ where $l = \mu,e$) with an overall 30\% branching ratio, 
and fully hadronic modes ($t\bar{t} \to (jj)(jj)b\bar{b}$) with an overall branching ratio 
of 46\%.

Semi-leptonic modes present the best signal-to-background ratio, with the dominant fake background 
being $W+nj$.
An analysis of KK modes for the LHC by Agashe \textit{et al.},~\cite{Agashe:2006hk} found an overall 
efficiency of $\approx\!1\%$, after taking into account branching fraction, cuts and $b$-tagging efficiency.
Their analysis included a 20\% efficiency for the $b$-tag.
They found the reducible background to be small compared to the irreducible QCD background.

In a separate analysis, Kaplan \textit{et al.}~\cite{Kaplan:2008ie} considered a method of top 
tagging for high $p_{T}$ hadronic top decays that suggests better efficiencies and does not incorporate 
a $b$-tag.
Their analysis using Pythia and a detailed systematic examination of top decays suggests a tagging 
efficiency of approximately 35\% to 45\% for top jets with transverse momentum between 600 and 
1400~GeV.
For fully hadronic $t\bar{t}$ events, this technique could result in an efficiency $>\!10\%$ for 
$\approx\!46\%$ of events, and a rejection against light jets $>\!10^4$.
Very specifically, the dijet reducible background is shown to be reduced to the same level as the QCD 
$t\bar{t}$ background, where a resonant peak may be visible.

Almeida \textit{et al.}~\cite{Almeida:2010pa} found that using only a cut on the jet mass
resulted in a single top jet identification efficiency of between 34\% and 58\% 
with a rejection factor of around 30.
By applying further cuts on the jet structure, they were able to achieve a rejection factor of about 5000 for a 
top identification efficiency of $\approx\!21\%$.
This is in general agreement with the Kaplan study in that 
reasonable top jet tagging efficiencies 
can be expected at the LHC without significant light jet contamination.

In our analysis, we assume the method used by Kaplan \textit{et al.} and employ an efficiency of 16\% 
(40\% for each top jet - the midpoint of the expected range) for the fully hadronic decay mode, with a 
rejection of dijets by a factor of $10^4$.

\subsection{Tagging Summary}

We summarize the tagging efficiencies we use in Table~\ref{table:tag}. 
For $\tau$ and $t$ decays, we only consider the fully hadronic decay channels. 
The efficiencies are used to determine the statistical uncertainties we give in our results.
For ratios of cross sections we assume that experimentalists have properly taken into account
efficiencies to extract the appropriate cross section.

\begin{table}[tbh]
\begin{center}
\caption[ Table of tagging efficiencies.] 
{\setlength{\baselineskip}{0.5cm} Summary of the overall efficiencies used for 
estimating the number of events observed of the given fermion species.
The overall efficiency is the tagging efficiency for the observed fermion decay mode times
the BR to that final state.}
\vspace{2mm}
\begin{tabular}{ l | c | c }
\textbf{Channel} & \textbf{Overall $\epsilon_{f}$} & \textbf{Jet Rejection $\epsilon_{j}$}\\
\hline
$Z^\prime \rightarrow \mu^+\mu^-$ 	& 0.92 & n/a\\
$Z^\prime \rightarrow \tau^+\tau^-$ 	& 0.08 & $2.0\times10^{-6}$\\
$Z^\prime \rightarrow b\bar{b}$ 	& 0.36 & $2.5\times10^{-5}$\\
$Z^\prime \rightarrow t\bar{t}$ 		& 0.075 & $1.0\times10^{-4}$\\
\end{tabular}
\label{table:tag}
\end{center}
\end{table}

\subsection{Kinematic Cuts}

\begin{figure}[htbp]
   \begin{center}
   \leavevmode
   \mbox{}
   \epsfxsize=8.5cm
   \epsffile{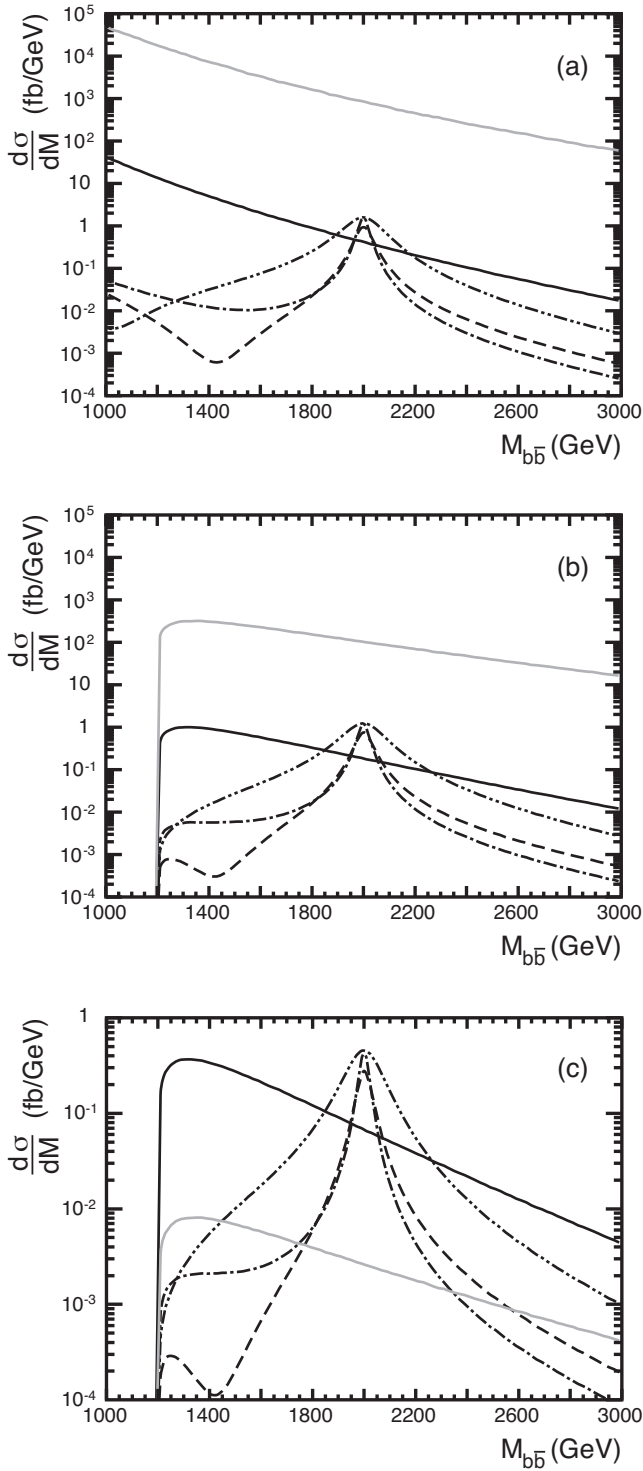}
   \end{center}
   \vspace{-8mm}
   \caption[ Invariant Mass distribution of $b$-quark distribution.]
{\setlength{\baselineskip}{0.5cm} 
$b\bar{b}$ invariant mass distributions for 
QCD $b\bar{b}$ background (solid, dark), light dijet background (solid, grey), and 
a $Z^{\prime}$ with $M_{Z^{\prime}}=2.0$~TeV for the LRM (dot-dash), UUM (dot-dot-dash), 
and LH (dashed) models.
(a) Only includes detector acceptance cuts of $p_T>20$~GeV and $|\eta|<2.5$.
(b) Also includes $p_T>0.3M_{Z^{\prime}}$ cut as described in the text but before including
tagging efficiencies. 
(c) Includes the application of cuts,  tagging efficiencies, 
and fake rejection rates given in Table~\ref{table:tag}. }
\label{fig:dsdm}
\end{figure}

The values for the fake rates and efficiencies assume the ideal case of perfect alignment within 
the inner detector and no pile-up.
Since tracking is only available within $|\eta|<2.5$, we do not include events in which one 
or both fermion tracks lie outside this region,
and, unless otherwise stated, we require a minimum $p_{T}>20$~GeV as needed for flavour tagging.
Figure~\ref{fig:dsdm}  
(a) shows the invariant mass distribution in the $b\bar{b}$ final state for several models, 
the QCD $b\bar{b}$ backgrounds, and the light dijet backgrounds.  

For hadronic decays of $\tau$, $b$ and $t$ fermions, we found that a $p_{T}>0.3 M_{Z{^\prime}}$ cut 
on the reconstructed momentum of the hadronic jets effectively reduced both the irreducible and 
dijet backgrounds as compared to the signal.
The events from the decay of a $Z^{\prime}$ tend towards a harder $p_T$ distribution than the 
QCD backgrounds 
so that a larger percentage of $Z^{\prime}$ events pass this cut than the 
QCD backgrounds.
Figure~\ref{fig:dsdm} (b) shows the signal for several $Z^{\prime}$ models, 
QCD $b\bar{b}$ background, and the light dijet background after applying this strong $p_T$ cut.
This cut was also applied to muon final states when calculating ratios of cross sections, 
for consistency.
For determination of the discovery limits we only apply a cut of $p_{T}>20$~GeV to the final state 
muons.

Fig.~\ref{fig:dsdm} (c) shows the reduction of the light dijet background after taking into account 
detector efficiencies and fake rates.
Figure~\ref{fig:dsdm} (c) shows that the application of appropriate flavour tagging algorithms 
reduces both the QCD $b\bar{b}$ and the light dijet backgrounds to a level where a 
meaningful measurement should be possible.

As a final kinematic cut, we include only events within an appropriate invariant mass window 
around the resonance mass to improve the signal to background. 
Unless otherwise stated,
we take this to be $|M_{Z^{/prime}} - M_{f\bar{f}}| < 2.5 \Gamma_{Z^{/prime}}$.
This restricts the background events to the kinematic
region directly under the resonance peak.  However, this introduces 
an additional experimental uncertainty due to detector resolution, 
which smears out the resonance peak, effectively reducing the number of
signal events in the peak.  Because we 
use ratios of cross section measurements 
into $b\bar{b}$, $t\bar{t}$, $\mu^+ \mu^-$, and 
$\tau^+ \tau^-$ final states, which have different energy resolutions,
the numerical value of the observables will shift.
Thus, it will be important to understand detector
resolution to accurately extract the underlying cross sections and coupling dependence. 
In addition, the
reduction of the measured signal compared to background will increase the experimental errors. 

To properly account for detector
resolution requires a realistic, detailed simulation for the specific particle 
identification algorithms being used. As discussed above, this subject is evolving rapidly.  
In addition, experimentalists 
are constantly improving their understanding of the energy calibration of the LHC detectors.
To gauge the importance of detector resolution, 
we use estimates from recent detector studies and include them 
by applying Gaussian smearing to the final state momentum. 
Studies by  the ATLAS collaboration expect 
between 3\% and 5\% energy resolution
for TeV scale hadronic jets \cite{ATLAS:2008zzm} 
(although the current ATLAS jet calibration gives 5\% resolution \cite{atlas-conf-pub-2010-054}).
This results in a resolution of $\sim\!3.5\%$ for $M_{b\bar{b}}$. 
This broadens the resonance, reducing the number of measured signal events within 
 the same invariant mass window. Signal significance may therefore
be significantly reduced for narrow resonances.

To see how detector resolution affects measurements, we
include results for the ideal case of no smearing and for a more realistic case of 5\% 
on $b$ and $t$ final states and 3\% on muon final states.  
For the $b\bar{b}$ case
we found that for the chosen mass window, as expected, detector
resolution has 
the greatest effect for the narrowest resonances (e.g. the $Z^\prime_{\psi}$)
with virtually no effect on the broadest states (e.g. the $Z^\prime_{UUM}$).

The situation for the $t\bar{t}$ final state is less conclusive for the following reason.  
A recent ATLAS study \cite{Aad:2009wy}
gives a resolution of 4.6\% for the $t\bar{t}$ invariant mass
distribution of a narrow resonance.  Using the same procedure used to gauge the importance of 
detector resolution for the $b\bar{b}$ channel
we find that a 
usable signal can be measured in the $t\bar{t}$
channel for all models considered in this paper except for possibly $Z_\eta$ and $Z_\psi$.  
This is consistent with the findings of Barger, Han and Walker \cite{Barger:2006hm}.
However, another ATLAS study \cite{atl-phys-pub-2010-008}
gives a resolution of 9-10\% for a 1~TeV $Z^{/prime}$ with width $\Gamma_{Z^{/prime}}/M_{Z^{/prime}}=3.3\%$. 
The range of these expectations demonstrates
the difficulty in trying to predict the detector resolution for these measurements. 

Reconstruction of $\tau^+\tau^-$ final states are complicated due to the 
missing energy from the neutrinos in the $\tau$ decays. LHC studies of $\tau^+\tau^-$ final 
states with respect to Higgs searches have found a resolution on the reconstructed Higgs mass of 
$M_H \sim 10 \% $\cite{Aad:2009wy,Leney:2008di,Elagin:2010aw}. We do not incorporate this mass 
resolution in our results, but will refer to it when discussing our results for $\tau^+\tau^-$ final states.

Thus, while we do estimate the effects that detector energy resolution will have on the 
precision of our measurements, past experience shows that
the experimentalists eventually exceed initial expectations.
In addition to this, the reduced signal to background caused by detector resolution can be mitigated to 
some extent by
higher luminosities and better identification efficiencies, which would improve the statistics.

\section{Results}

\subsection{Discovery Limits}

Leptonic final states offer the cleanest channel for the discovery of extra neutral gauge bosons
due to the low backgrounds and clean identification
\cite{Godfrey:1994qk, Godfrey:2002tna, Capstick:1987uc, Thompson:2008tz, Kang:2004bz}.
A very small number of dilepton events clustered in one or two bins of the invariant mass distribution
would be taken as an obvious signal for new physics.
To quantify this we consider two opposite sign leptons and impose kinematic cuts
of $|\eta_l|<2.5$ and $p_{T_l}>20$~GeV  to reflect detector acceptance.
The criteria used was 5 events in the $\mu^+\mu^-$ channel with a signal-over-background of at 
least 5
in an invariant mass window within $\pm 1$ bins of the resonance peak with the bin size as 
defined by $\Delta M = 24 (0.625 M + M^2 + 0.0056)^{1/2}$~GeV where $M$ is given
in TeV~\cite{atlas}.

\begin{figure}[t]
   \begin{center}
   \leavevmode
   \mbox{}
   \epsfxsize=8.5cm
   \epsffile{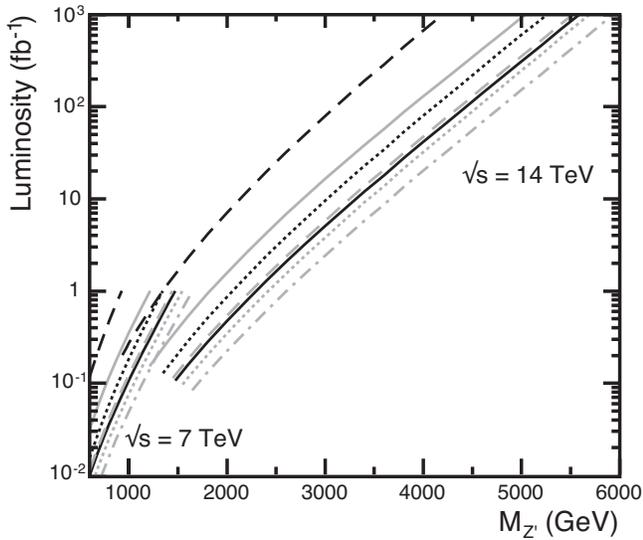}
   \end{center}
   \vspace{-8mm}
   \caption[ Expected luminosity for discovery of a $Z^{\prime}$ for each model considered.]
{\setlength{\baselineskip}{0.5cm} Luminosity required for $Z^{\prime}$  
discovery as a function of $M_{Z^{\prime}}$ based on   
the observation of 5 $\mu^{+}\mu^{-}$ events within the 
invariant mass window described in the text.  The sets of curves are for 
an LHC centre of mass energy of  $\sqrt{s}=7$~TeV and  $\sqrt{s}=14$~TeV.
>From left to right, the models are ETC (dashed, dark), TC2 (solid, grey), 
UUM (dotted, dark), AFSLH (dashed, grey), SLH (solid, dark), 
SSM (dotted, grey) and LH (dot-dash, grey).}
\label{fig:lum_m}
\end{figure}

The integrated luminosity required to discover a $Z^{\prime}$ of a given mass in the dimuon channel 
is shown in Fig.~\ref{fig:lum_m}.
We show curves for $pp$ collisions with $\sqrt{s}=7$~TeV corresponding to the 2010-2011 LHC run 
and for $\sqrt{s}=14$~TeV which corresponds to the LHC design energy.
 
\begin{figure}[t]
   \begin{center}
   \leavevmode
   \mbox{}
   \epsfxsize=8.5cm
   \epsffile{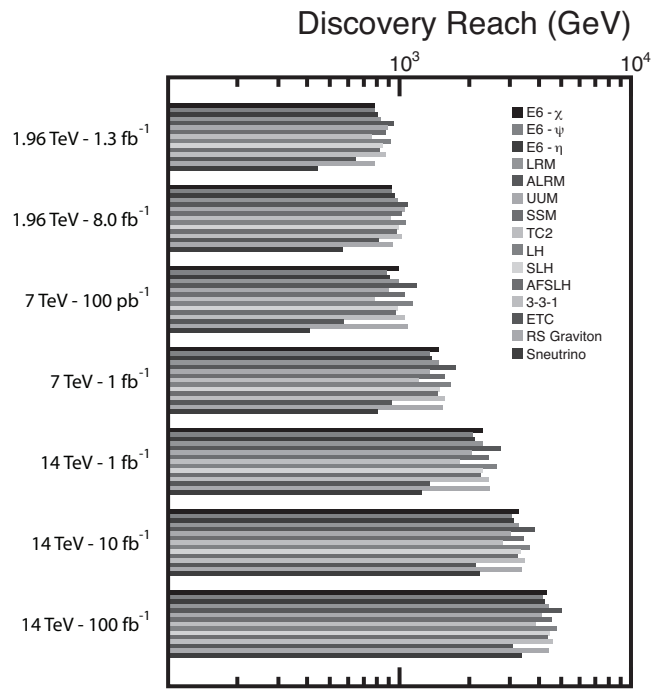}
   \end{center}
   \vspace{-7mm}
   \caption{Discovery reach at benchmark luminosities for Tevatron and LHC (both early and design) 
energies.}
\vspace{-4mm}
\label{fig:dl}
\end{figure}

In Fig.~\ref{fig:dl} we show discovery limits for the various models for several LHC 
benchmark energies and luminosities and compare them to discovery limits for the 
Fermilab $p\bar{p}$ collider.
For the Tevatron we assume two cases; 1.3~fb$^{-1}$ of integrated luminosity, as in 
Ref.~\cite{Aaltonen:2008vx}, and 8~fb$^{-1}$ to estimate the reach for the full expected luminosity.
We used similar detector acceptance and cuts as in Ref.~\cite{Aaltonen:2008vx}: 
we impose a kinematic cut of $p_{T}>25$~GeV and consider events within two regions of
pseudorapidity - where both leptons satisfy $|\eta|<1.1$, and where one lepton satisfies 
$|\eta|<1.1$ and the other satisfies $1.2<|\eta|<2.0$.
As well, we consider only events within an invariant mass window of 
$|M_{Z^{\prime}}-M_{l^+l^-}| = \pm10\% M_{Z^{\prime}}$.
We use the discovery criteria of 5 observed dilepton events as with the LHC study.
We note that this differs from that used by the CDF collaboration to obtain the current direct 
limits in Refs.
\cite{Aaltonen:2008vx} and~\cite{Aaltonen:2008ah}.
Figure~\ref{fig:dl} shows that the 2010-2011 LHC run will be able to roughly double 
the discovery reach of the Tevatron.

In addition, we considered the possibility that a $Z^{\prime}$ with a preferential coupling 
to third generation fermions may be discovered first in the $\tau^+ \tau^-$ final state.
For the $\tau$ hadronic decay mode, none of the models we considered would be
first visible in the $\tau$ final state for a value of mixing angle, $\phi$, that would give rise to 
a narrow resonance.
At extreme values of the mixing angles in models without generation universality, the $Z^{\prime}$ 
begins to decouple from the first two generation fermions, reducing the $q\bar{q}-Z^{\prime}$ 
coupling, and hence the production of the $Z^{\prime}$.
For these models, a $Z^{\prime}$ that would be first observed in the $\tau^+\tau^-$ final state
would have a width larger than 10\% of the $Z^{\prime}$ mass, resulting in a very broad peak,
unlikely to be observed.
However, as described by Holdom~\cite{Holdom:2008xx}, 
if other $\tau$ decay modes could be used, it may be possible to first observe a $Z^{\prime}$ in the
$\tau^+\tau^-$ final state for some regions of parameter space.

Constraints from electroweak precision data are more stringent than the discovery limits 
from the Tevatron for some models, as seen in 
Refs.~\cite{Erler:2009jh,delAguila:2010mx,Chivukula:2002ry}. 
However, it is clear that the 7~TeV LHC run will improve the limits for models with 
universal couplings, and the full 14~TeV run should improve the limits for non-universal 
models for reasonable values of the mixing angle, $\phi$.

\subsection{Model Discrimination using $t\bar{t}$ and $b\bar{b}$ to $\mu^+ \mu^-$ Production Ratios}

The primary goal of this paper is to explore the use of third generation fermions to distinguish
between models of extra neutral gauge bosons.  We start with ratios of 
$t\bar{t}$ and $b\bar{b}$ to $\mu^+ \mu^-$ cross sections, 
and expand on our previous study~\cite{Godfrey:2008vf}.
We are particularly interested in models with non-universal couplings - specifically,
the UUM, ETC and TC2 models - as 
the $R_{t/\mu}$ and $R_{b/\mu}$ ratios defined below produce results that are quite 
distinctive from the models considered in Ref.~\cite{Godfrey:2008vf}.

$R_{t/\mu}$ and $R_{b/\mu}$ are defined by:
\begin{eqnarray}
\label{eqn:ratio_t}
R_{t/\mu} \equiv \frac{\sigma(pp \to Z^{\prime} \to t\bar{t})}{\sigma(pp \to Z^{\prime} \to \mu^{+}\mu^{-})} 
\approx \frac{3K_{t}\left(L_{t}^{2} + R_{t}^{2}\right)}{\left(L_{\mu}^{2} + R_{\mu}^{2}\right)} \\
\label{eqn:ratio_b}
R_{b/\mu} \equiv \frac{\sigma(pp \to Z^{\prime} \to b\bar{b})}{\sigma(pp \to Z^{\prime} \to \mu^{+}\mu^{-})} 
\approx \frac{3K_{b}\left(L_{b}^{2} + R_{b}^{2}\right)}{\left(L_{\mu}^{2} + R_{\mu}^{2}\right)} ,
\end{eqnarray}
where $L_{f}$ and $R_{f}$ are the left and right handed fermion couplings to the $Z^{\prime}$ and 
the $K$ factors incorporate the QCD 
and QED  NLO correction factors\cite{KubarAndre:1978uy,Kataev:1992dg}.
The use of the $R_{t/\mu}$ and $R_{b/\mu}$ ratios has the benefit of reducing the contributions 
from uncertainties in the parton distribution functions as the initial state $q\bar{q}$ 
couplings and pdf's cancel in the ratio.

To obtain our results, we assumed that a $Z^{\prime}$ has been discovered and its mass and width 
measured~\cite{Langacker:1984dc,Barger:1980ix, Godfrey:1994qk, Capstick:1987uc} such that the appropriate $M_{Q\bar{Q}}$ and $p_{T}$ cuts 
described above can be applied.
We calculated the expected number of events and statistical errors for signal plus background 
for a given integrated luminosity and 
particle identification efficiencies, $\epsilon_{\mu^+\mu^-}$,  
$\epsilon_{b\bar{b}}$, and $\epsilon_{t\bar{t}}$ from Table~\ref{table:tag}.
The expected number of SM QCD background events were 
subtracted from the 
total events to give the predicted number of signal events.
We did not include systematic uncertainties arising from uncertainties 
in the luminosity and identification efficiencies.

\begin{figure}[t]
   \begin{center}
   \leavevmode
   \mbox{}
   \hspace{-8mm}
   \epsfxsize=8.5cm
   \epsffile{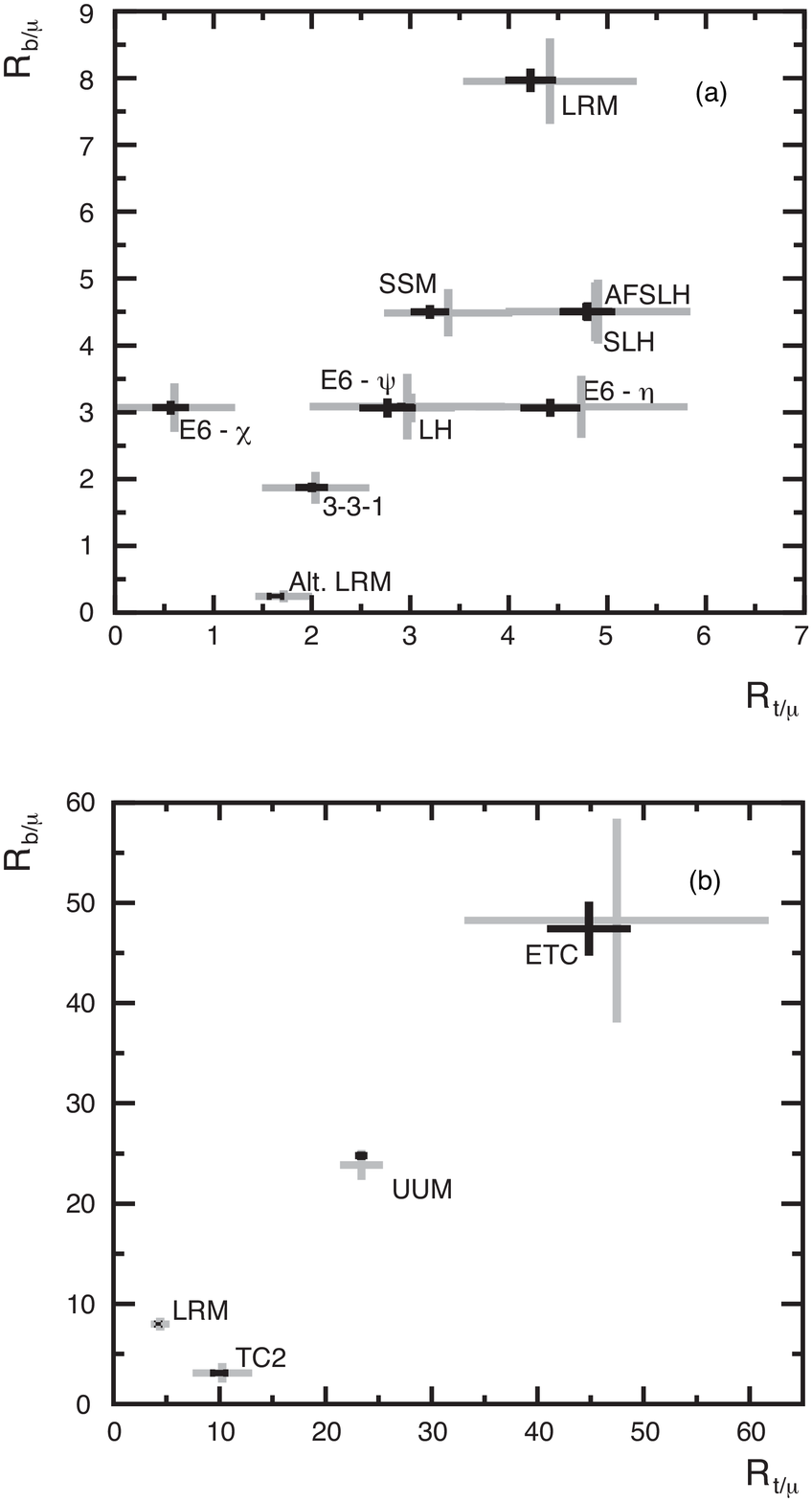}
   \end{center}
   \vspace{-10mm}
   \caption[ Plot showing the distribution of R_{t/\mu} versus R_{b/\mu} points for each models.]
{\setlength{\baselineskip}{0.5cm} Measurements of the ratio of $t\bar{t}$ and $b\bar{b}$ 
to $\mu^{+}\mu^{-}$ cross sections within $2.5\Gamma$ of $M_{Z^{\prime}}$. 
Black bars correspond to expected 1$\sigma$ statistical uncertainties for 
$M_{Z^{\prime}} = 1.5$~TeV and grey bars to $M_{Z^{\prime}} = 2.5$~TeV.
(b)  Expands the scales 
to include models with generation and family dependent couplings.}
\label{fig:ratio_mass}
\end{figure}

Our results for  $R_{b/\mu}$ and $R_{t/\mu}$ are shown in Fig.~\ref{fig:ratio_mass} (a) and (b), where 
Fig.~\ref{fig:ratio_mass} (b) expands the view to show 
the distinctive measurements that would be observed for models with preferential couplings.
We show  $1\sigma$ statistical errors based on an integrated luminosity of $L=100$~fb$^{-1}$.
The measurements for models with preferential couplings and mixing parameters that 
result in large couplings for third generation quarks are quite distinctive, 
making this a possible hallmark measurement for these models.

\begin{figure}[t]
   \begin{center}
   \leavevmode
   \mbox{}
   \hspace{-8mm}
   \epsfxsize=8.5cm
   \epsffile{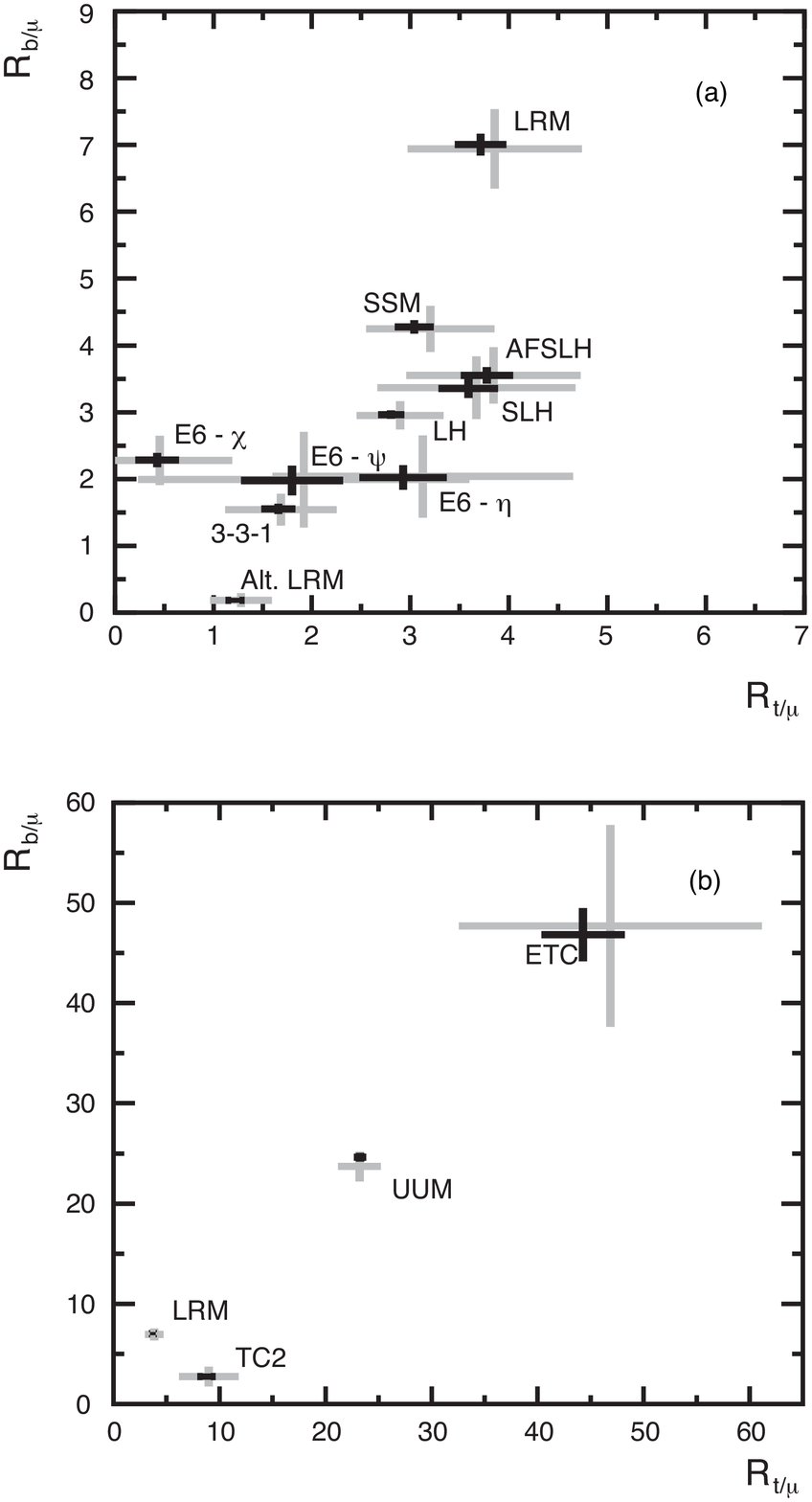}
   \end{center}
   \vspace{-10mm}
   \caption[ Plot showing the distribution of R_{t/\mu} versus R_{b/\mu} points for each models.]
{\setlength{\baselineskip}{0.5cm} Measurements of the ratio of $t\bar{t}$ and $b\bar{b}$ 
to $\mu^{+}\mu^{-}$ cross sections within $2.5\Gamma$ of $M_{Z^{\prime}}$ including detector 
resolution of 5\% for $b$ and $t$ final states, and 3\% for muons. 
The labelling is as in Fig.~\ref{fig:ratio_mass}.}
\label{fig:ratio_smeared}
\end{figure}

In Fig.~\ref{fig:ratio_smeared} we show $R_{b/\mu}$ and $R_{t/\mu}$ including a detector 
resolution of 5\% for $b$ and $t$ 
final states, and 3\% for muons.  Comparing Fig.~\ref{fig:ratio_smeared}  
with Fig.~\ref{fig:ratio_mass}, it is clear that the effect of resolution is small for broad models such 
as ETC, TC2 and UUM.
On the other hand, the effect is much more significant for the narrow models such 
as the E6, SLH, and AFSLH models, affecting both the values and expected statistical 
uncertainty. There are two approaches 
to account for resolution effects in these measurements.
The experimentalists can try to disentangle resolution 
effects from the underlying cross section to give the ``theoretical'' cross section or they can 
compare to Monte Carlo simulations that include detector resolution.   
In any case, the sensitivity of 
measurements for narrow models to detector resolution will lead to 
systematic uncertainties in the measurements 
that need to be taken into account.

\begin{figure}[t]
   \begin{center}
   \leavevmode
   \mbox{}
   \hspace{-8mm}
   \epsfxsize=8.5cm
   \epsffile{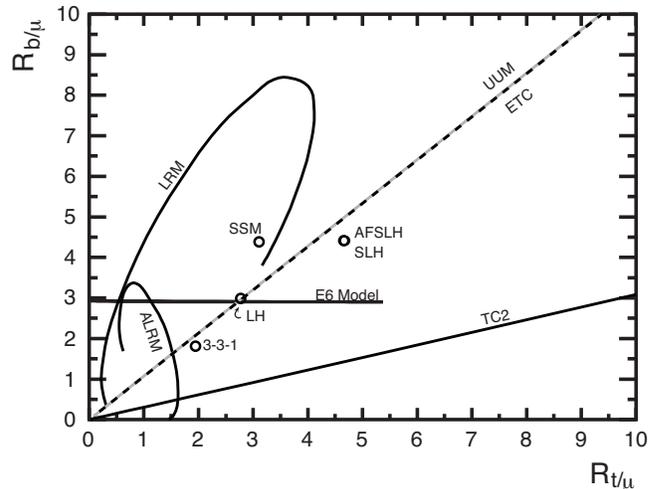}
   \end{center}
   \vspace{-10mm}
   \caption[ Effect of varying free parameters for each model.]
{\setlength{\baselineskip}{0.5cm} 
$R_{b/\mu}$ vs $R_{t/\mu}$ spanning the model parameter range as given in the text,
for the models labelled in the figure.}
\label{fig:ratio_param}
\end{figure}

For some models, including models with non-universal couplings, 
the $E_{6}$ model, and the LR Symmetric models, the 
predictions for $R_{b/\mu}$ and $R_{t/\mu}$
are dependent on the mixing angle between subgroups in the model.
This is illustrated in Fig.~\ref{fig:ratio_param} which plots the $R_{t/\mu}-R_{b/\mu}$ 
ratios while varying the mixing parameters.  
To obtain these curves, a $Z^{/prime}$ mass of 1.5~TeV 
was used, although the results are not very sensitive
to $M_{Z^{/prime}}$, and the following ranges for mixing angles were used: 
for the LR Symmetric model, both with standard and alternate isospin assignments, the 
mixing parameter is constrained by $0.55 \leq \left(g_{R}/g_{L}\right) \leq 1$~\cite{Chang:1984uy};
for the UUM, $\phi$ is constrained by $0.22 \leq \sin\phi \leq 0.99$~\cite{Barger:1989bb};
and for the remaining models no specific limits could be found in the literature 
that were not directly tied to the mass of the $Z^{\prime}$.  
Depending on the mixing parameter, the predictions for some models overlap in the 
$R_{t/\mu}-R_{b/\mu}$ space as shown in Fig.~\ref{fig:ratio_param}.
Consequently, other measurements will be needed to distinguish between 
models for these values of the model parameters.

It is important to note that the region of overlap in the $R_{t/\mu}-R_{b/\mu}$ plane between 
models with and without universal couplings occurs for parameter values 
that are not of particular 
interest.  In the case of the UUM model, the overlap occurs for parameter values 
where leptons have 
preferential couplings and in the case of the ETC and TC2 models the overlap occurs when
the first two generations of fermions have preferential couplings. 
Since these models are constructed such that the top quark plays a role in EWSB, 
one would not expect their mixing angles to take values in the overlap region.

A final observation is that 
the UUM and ETC models are indistinguishable using measurements of $R_{t/\mu}$ and $R_{b/\mu}$
for any value of mixing parameter. However,  in this case, 
the ratio of tau to muon events at the LHC will discriminate generation 
dependent couplings.  This is examined in the next subsection.

\subsection{Extracting Mixing Parameters Using $R_{\tau/\mu}$}

Measuring generation universality will be an important step in distinguishing or ruling out TC2 and 
ETC type models.
The simplest and cleanest way to measure the level of universality is to determine the ratio 
of the $Z^{\prime}$ decays to $\tau$-leptons and to muons or electrons which can be found by
measuring $\sigma(pp\to Z^{\prime} \to \tau^+\tau^-)/\sigma(pp\to Z^{\prime} \to \mu^+\mu^-)$.

We also considered a measurement of the ratio of $t$- to $c$-quark cross sections but found that 
such a measurement does not appear promising due to low tagging efficiencies of the charm 
quark, and indistinguishability of the charm quark from the light jet backgrounds.

\begin{figure}[t]
   \begin{center}
   \leavevmode
   \mbox{}
   \epsfxsize=8.5cm
   \epsffile{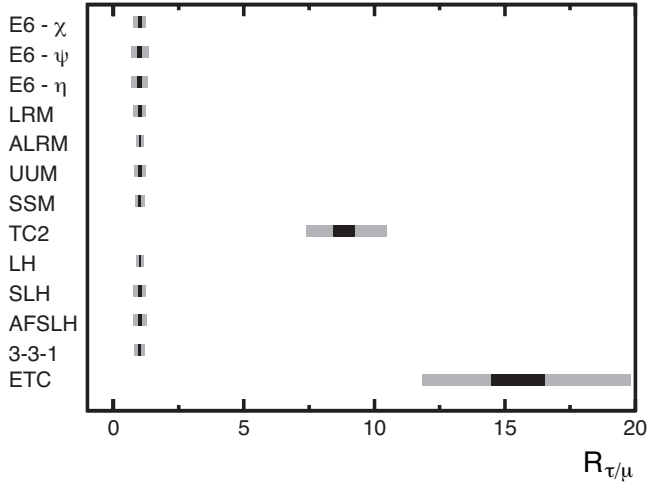}
   \end{center}
   \vspace{-4mm}
   \caption{$R_{\tau/\mu}$ for models with universal and non-universal fermion couplings. 
The ratios are obtained for tau and muon events satisfying $|M_{f\bar{f}}-M_{Z^{\prime}}|< 2.5\Gamma_{Z^\prime}$,
along with the other cuts described in the text.  
The dark bars are for $M_{Z^{\prime}}=1.5$~TeV and the grey bars for $M_{Z^{\prime}}=2.5$~TeV
(for $\sin\phi_{TC2} = 0.5$ and $\sin\phi_{ETC} = 0.9$).}
\label{fig:lepton_gen}
\end{figure}

\begin{figure}[t]
   \begin{center}
   \leavevmode
   \mbox{}
   \epsfxsize=8.5cm
   \epsffile{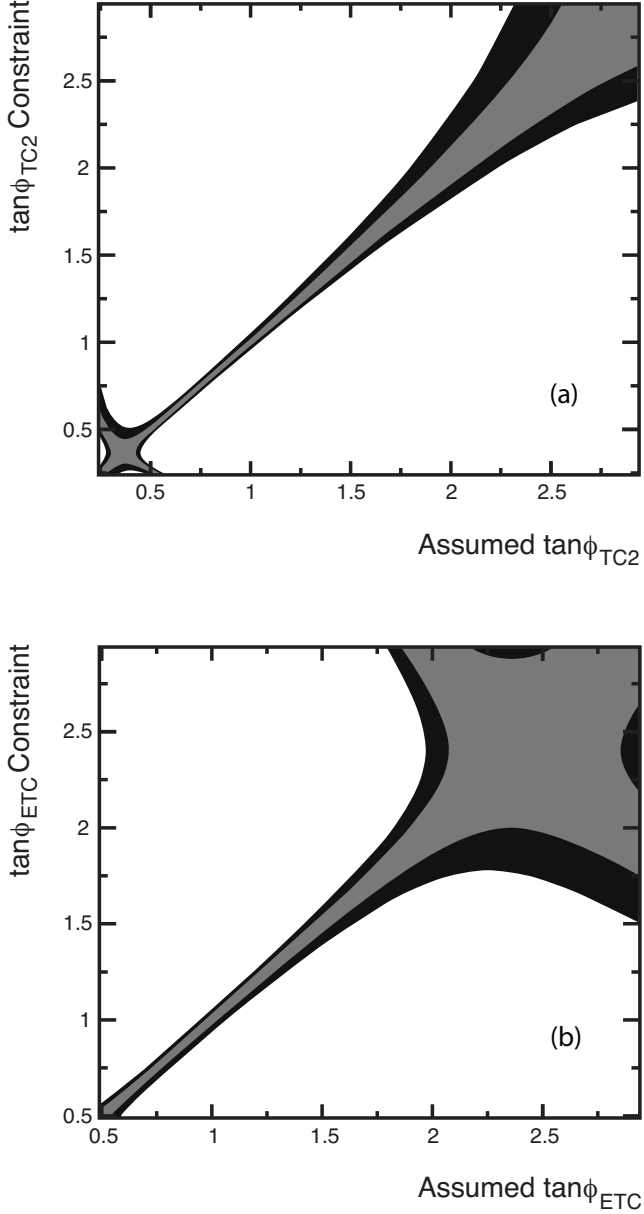}
   \end{center}
   \vspace{-6mm}
   \caption[ $\chi^2$ analysis of mixing parameter ]
{\setlength{\baselineskip}{0.5cm} 
LHC measurement capability of the mixing parameter using $R_{\tau/\mu}$ for (a) TC2
and (b) ETC.  The horizontal axis corresponds to the input parameter value and the verticle
axis corresponds to the extracted value with 1- (grey) and 2- (black) $\sigma$ limits. $M_{Z^{\prime}} = 
1.5$~TeV and an integrated luminosity of 100~fb$^{-1}$ was assumed in these plots with the following
kinematic cuts; $|\eta|<2.5$, 
$|M_{l^+l^-} - M_{Z^{\prime}}|<2.5\Gamma_{Z^{\prime}}$, $p_T< 0.3 M_{Z^\prime}$.}
\label{fig:theta_comb}
\end{figure}

We therefore restrict ourselves to measurement of the ratio
$R_{\tau/\mu}=\sigma(pp\to Z^{\prime} \to \tau^{+}\tau^{-})/\sigma (pp \to Z^{\prime} \to \mu^{+}\mu^{-})$, 
shown in Fig.~\ref{fig:lepton_gen}.  To obtain these results we imposed a cut on the invariant
mass of the final state fermions of $|M_{Z^{\prime}}-M_{f\bar{f}}|< 2.5\Gamma_{Z^{\prime}}$ 
in addition to imposing a requirement of $p_{T_f}>0.3 M_{Z^{\prime}}$ and $|\eta_f|<2.5$.  
These results and those shown in Fig.~\ref{fig:theta_comb}
only include the statistical errors for a semi-idealized detector with perfect energy resolution. 
We will comment on this below.

It is clear from Fig.~\ref{fig:lepton_gen} that models with generation universality will yield
measurements of $R_{\tau/\mu}\simeq 1$ with reasonable precision.  In contrast, 
models with generation dependent couplings 
show a large, measurable
variation from unity.  The dependence of $R_{\tau/\mu}$ on the mixing angle between the 
gauge groups of the theory are given by:
\begin{equation}
\frac{\Gamma(Z^{\prime}_{ETC} \to \tau^{+}\tau^{-})}{\Gamma(Z^{\prime}_{ETC} \to \mu^{+}\mu^{-})} \propto 
\tan^{4}\phi_{ETC}
\end{equation}
and
\begin{equation}
\frac{\Gamma(Z^{\prime}_{TC2} \to \tau^{+}\tau^{-})}{\Gamma(Z^{\prime}_{TC2} \to \mu^{+}\mu^{-})} \propto 
\cot^{4}\phi_{TC2}
\end{equation}

Given the fundamental nature of the gauge group mixing angle, measuring its precise value 
would be vital input into constructing the Lagrangian of the underlying theory.  
In Fig.~\ref{fig:theta_comb} we show how well such a measurement can be made for the TC2 and ETC 
models using $R_{\tau/\mu}$ assuming $M_{Z^{\prime}}=1.5$~TeV and $L=100$~fb$^{-1}$
for the semi-idealized detector.  
In these plots the $x$-axis corresponds to the assumed value of the mixing parameter and 
the $y$-axis corresponds to the measured value of the mixing parameter with the spread in the 
vertical direction corresponding to 1- and 2-$\sigma$ limits of $R_{\tau/\mu}$ for the
input parameter value and measured value.
The parameter range corresponds to the range where the $Z^{\prime}$ width is less than 10\% of the 
$Z^{\prime}$ mass.
As before, we include the
backgrounds in estimating the statistical errors and impose the same kinematic cuts as before.
These limits could be further constrained by including more observables, such as
$R_{b/\mu}$ and $R_{t/\mu}$ into the fit.

Figure~\ref{fig:lepton_gen} indicates that, for typical parameter values, the non-universal models
predict values for $R_{\tau/\mu}$ that
are very distinct from the value of $R_{\tau/\mu}\simeq 1$
expected for models with generation universality.  
As pointed out, these results do not include finite detector resolution and
in addition, measurements of $\tau^+\tau^-$ final states have the additional 
complication of missing energy due to neutrinos in the $\tau$ decays. 
However, the non-universal $Z^{/prime}$'s have large decay widths so that including the measurement
resolution \cite{Aad:2009wy,Leney:2008di,Elagin:2010aw}
will only have a small effect on the results for non-universal $Z^{/prime}$'s.  
On the other hand, some of the 
generation universal $Z^{/prime}$ models are relatively narrow so that the resonances will be smeared 
out in the $\tau^+\tau^-$ final state
and some care will have to be taken in choosing appropriate invariant mass windows. 
We note that both 
ATLAS \cite{atl-phys-pub-2011-010} and CMS \cite{Chatrchyan:2011nv} have recently measured
$\sigma (pp\to Z^0) \times {\cal B}(Z^0\to \tau^+ \tau^-)$ with errors of 
roughly  20\% and 10\% respectively in the $\tau_h \tau_\ell$ modes which bodes well for 
a measurement of $R_{\tau/\mu}$.

We conclude that, at worse, a very crude measurement of $R_{\tau/\mu}$ could signal generation 
non-universality. 
The mixing parameter measurement capability shown in Fig.~\ref{fig:theta_comb} is an 
idealization but, given the recent ATLAS and CMS measurements of $Z^0\to \tau^+ \tau^-$, we
are optimistic that a measurement will be possible to constrain the relevant mixing angle.

\subsection{Forward-Backward Asymmetry with Heavy Quark Final States}

The ability to identify $b$- and $t$-quarks at the LHC offers the possibility of 
using forward-backward asymmetries ($A_{FB}$) in heavy quark final states to 
assist in the determination of individual fermion couplings to a $Z^{\prime}$~\cite{Diener:2009ee}. 
As will be seen below, $A_{FB}$ has a different dependence on
the $Z^{\prime}$-fermion couplings than the $Z^{\prime}$ production cross section and $Z^{\prime}$ 
width.

Generally, a forward event is defined by the decay angle of the outgoing fermion 
relative to the direction of the interacting quark in the Drell-Yan annihilation.
For $pp$ collisions at the LHC, there is an ambiguity in determining the direction of the quark, 
where it is impossible to tell on an event-by-event basis whether the $Z^{\prime}$ is boosted in 
the direction of the quark or anti-quark.
Because the momentum distributions are harder for valence quarks than for sea anti-quarks, this 
ambiguity can be resolved to a certain extent by assuming that the $Z^{\prime}$ boost direction 
is the same as the quark direction~\cite{Dittmar:1996my}.
In the central region of $Z^{\prime}$ rapidity, the quark and anti-quark momenta are more evenly 
balanced, and the correct and incorrect assignments are canceled in an asymmetry measurement.

In a recent paper~\cite{Diener:2009ee}, we suggested a simple method of performing the 
forward-backward asymmetry measurement by using the direct pseudorapidity 
measurements of the final state particles. This method differs from the traditional definition given in 
\cite{Langacker:1984dc, Anderson:1992jz, delAguila:1993ym}.
It can be shown that a ``forward" event is one in which $|\eta_{f}|>|\eta_{\bar{f}}|$ in the lab frame, and 
vice-versa for a ``backward" event, assuming that the $Z^{\prime}$ is boosted by the quark as in 
the traditional definition.
Using these forms for forward and backward events, the forward-backward asymmetry of the 
signal is given by:
\begin{eqnarray}
A_{FB} &=& \frac{\displaystyle\int F(y) - B(y) dy}{\displaystyle\int F(y) + B(y) dy} \cr
& \sim & \displaystyle\left( \frac{{L_Q}^2 - {R_Q}^2}{{L_Q}^2 + {R_Q}^2}\displaystyle\right)
\left( { { \displaystyle\sum_q G_q^-({L_q}^2 - {R_q}^2 )} \over { 
\displaystyle\sum_q G_q^+( {L_q}^2 + {R_q}^2) } }\right)
\label{eq:afb}
\end{eqnarray}
where $F(y)$ is the number of forward events and $B(y)$ is the number of backward events for a given
$Z^{\prime}$ rapidity, $y$.
The approximation is a representation of the on-peak contribution to the $A_{FB}$ that more clearly 
indicates the coupling dependencies.
$L_f$ and $R_f$ are the left and right handed couplings of the $Z^{\prime}$ to the fermions, and 
$G^{\pm}_q$ are the integrated symmetric and anti-symmetric combinations of the parton distribution 
functions.

\begin{figure}[t]
\centering
\includegraphics[width=80mm,clip]{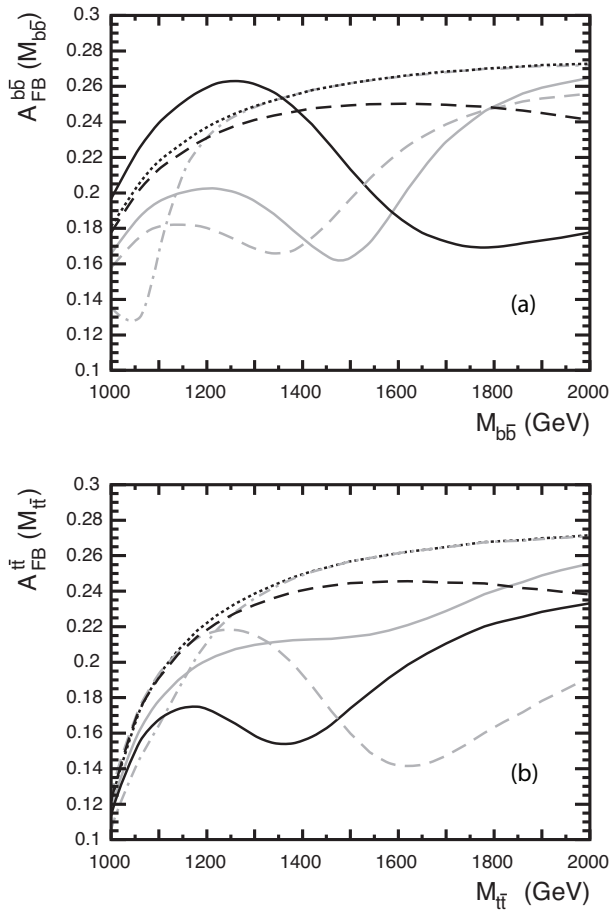} \\
\caption{ (a) $A_{FB}^{b\bar{b}}$ and (b) $A_{FB}^{t\bar{t}}$ of the $Z^{\prime}$ signal as a function of 
invariant mass for a $Z^{\prime}$ with $M_{Z^{\prime}} = 1.5$~TeV. Cuts employed include $p_{T}>0.3 
M_{Z^{\prime}}$ and $|\eta|<2.5$. These figures do not include backgrounds, and only examine the 
asymmetry distribution of $pp \rightarrow \gamma/Z^0/Z^{\prime} \rightarrow q\bar{q}$. 
Models follow the same format as Fig.~\ref{fig:lum_m}: ETC (dashed, dark), TC2 (solid, grey), 
UUM (dotted, dark), AFSLH (dashed, grey), SLH (solid, dark), 
SSM (dotted, grey) and LH (dot-dash, grey). The results are for the ``ideal'' case that doesn't 
take into account detector resolution.}
\vspace{-2mm}
\label{fig:afb_dist}
\end{figure}

\begin{figure}[t]
\centering
\includegraphics[width=85mm,clip]{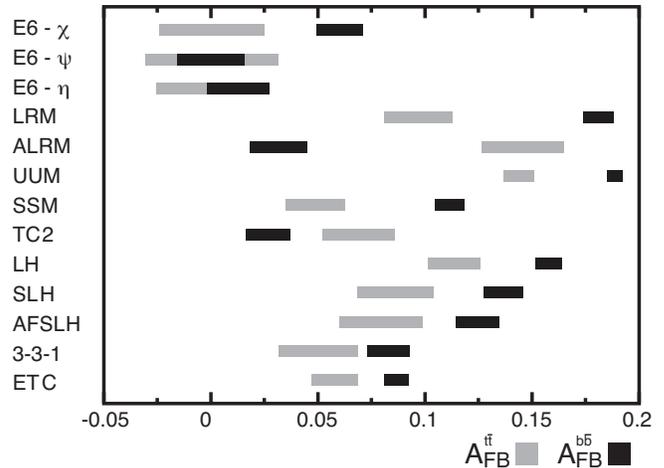} \\
\caption{$A_{FB}^{b\bar{b}}$ (black bars) and $A_{FB}^{t\bar{t}}$ (grey bars) for a $Z^{\prime}$ with a 
mass of 1.5~TeV.
Statistical errors include contributions from QCD backgrounds and light dijets 
assuming 100~fb$^{-1}$ luminosity.}
\label{fig:afb}
\end{figure}

\begin{figure}[t]
\centering
\includegraphics[width=85mm,clip]{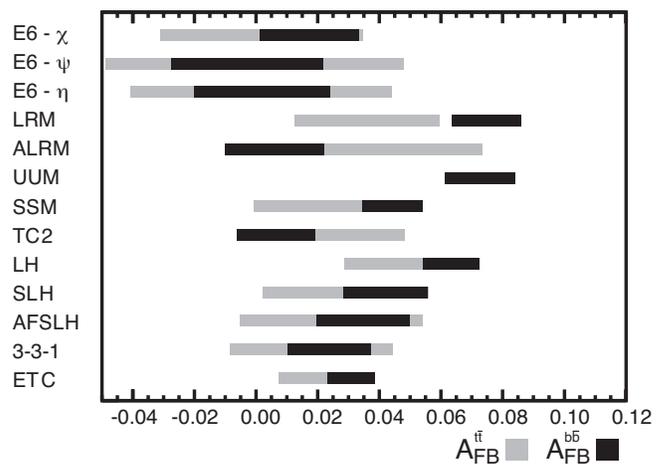} \\
\caption{$A_{FB}^{b\bar{b}}$ (black bars) and $A_{FB}^{t\bar{t}}$ (grey bars) including 
detector resolution as described in the text.  
Statistical errors include contributions from QCD backgrounds and light dijets 
assuming 300~fb$^{-1}$ luminosity.}
\label{fig:afb-smear}
\end{figure}

This method for finding the $A_{FB}$ has the advantage of being very straightforward and clean.
It simply relies on counting events with $|\eta_{f}|>|\eta_{\bar{f}}|$ and 
$|\eta_{f}|<|\eta_{\bar{f}}|$.
No calculation of the centre-of-mass scattering angle or $Z^{\prime}$ rapidity is required. 
In Fig.~\ref{fig:afb_dist}, we show the $A_{FB}$ distribution as a function of invariant mass 
of the $b\bar{b}$ and $t\bar{t}$ final states for several representative models with a $Z^{\prime}$ 
mass of 1.5~TeV. It should be noted that it is unlikely there will be sufficient statistics 
to make this measurement except on the $Z^{/prime}$ resonance.

As in all measurements involving third generation fermions, 
the challenge is extracting the events of interest from a large standard model 
background and accumulating sufficient statistics to make a meaningful measurement.
The QCD backgrounds for these measurements are forward-backward 
 symmetric at tree level, which should allow a heavy quark $A_{FB}$ to be sensitive to the 
 presence of a $Z^{\prime}$.  
Rather than subtracting out the backgrounds we include the totals of signal plus backgrounds in 
the forward and backward regions.  
Since the backgrounds are symmetric, they 
do not contribute to the numerator, but do for the denominator.
This has the effect of making the magnitude of the asymmetry 
smaller but also has the advantage of reducing the statistical errors.
The statistical errors include contributions from both QCD heavy quark backgrounds and the 
reduced light dijet background.

Figure~\ref{fig:afb} shows the expected results for $A_{FB}$ in the $b\bar{b}$ and $t\bar{t}$ 
channels for several models assuming $M_{Z^{\prime}}=1.5$~TeV, $L=100$~fb$^{-1}$,  and the 
same kinematic cuts described previously,
not including detector resolution effects.  
While large uncertainties are apparent for some models, a reasonable measurement can still be 
expected for most models, including the Left-Right Symmetric Model, 
the various Little Higgs models, and models with non-universal couplings.
Table~\ref{table:afb} gives the corresponding numerical values for $A_{FB}^{t\bar{t},b\bar{b}}$ 
and the
statistical uncertainties that can be expected for $100$~fb$^{-1}$ integrated luminosity.

Fig.~\ref{fig:afb-smear} shows similar results but this time including the detector resolution
as described earlier. 
In this figure we assumed an integrated luminosity of $300$~fb$^{-1}$ to improve the statistics.
Taking into account detector resolution both 
reduces the statistics and shifts the observed values for $A_{FB}$.
While $A_{FB}$ still has some resolving power with $b\bar{b}$ final states
it is not clear how useful the $t\bar{t}$ final states will be, due to low statistics.  
It is possible that if some effort were made to disentangle 
detector resolution from the underlying cross section these results could be improved.
Nevertheless,  we expect that $A_{FB}$ measurements would help constrain 
 the $Z^{\prime}$-fermion couplings as part of a global fit. 

\section{Summary}

Extra neutral gauge bosons are a hallmark of many models of physics beyond the standard model and
may be discovered early in the LHC program.
In this paper, we expanded 
upon our previous phenomenological study by focusing on measurements 
using third generation fermions and exploring models with non-universal fermion-$Z^{\prime}$ 
couplings.
We first gave an update of $Z^{\prime}$ discovery limits including non-universal coupling models 
and the LHC energy and planned luminosity for the 2010-2011 run.
We found that this run will roughly double the mass reach of the Fermilab Tevatron.

The main focus of this paper was to explore the usefulness of third generation fermions in 
studying extra neutral gauge bosons.
We found that it should be possible to measure the decays of a moderately heavy $Z^{\prime}$ 
to third 
generation quarks, depending on the capability of the experiments to reject against the light jet 
background.
Such measurements would prove to be very effective at distinguishing between different models of 
$Z^{\prime}$.  The measurement of the ratio of $\tau$ to $\mu$ decays 
of the $Z^{\prime}$ should be very effective in testing generation universality of the $Z^{\prime}$.
We also studied using the forward backward asymmetry of $b$- and $t$-quarks from $Z^{\prime}$ decays 
to distinguish models.  For a $Z^{\prime}$ of moderate mass these measurements could help distinguish 
between different models.  More importantly, they can contribute valuable input for measuring 
$Z^{\prime}$-fermion couplings, and should be an integral piece of a global fit with this goal.  

An important source of uncertainty in these measurements is the degradation of the heavy fermion
final state signals due to detector resolution. It appears to be manageable for $b\bar{b}$ and
$t\bar{t}$ final states for broad $Z^{/prime}$ resonances such as in the TC2 and ETC 
models, but becomes increasingly important for narrow resonances such as the E6 models.  
Now that ATLAS and CMS have demonstrated the ability to measure 
$\sigma(pp\to Z^0) \times {\cal B}(Z^0 \to \tau^+\tau^-)$ we are optimistic that
$R_{\tau/\mu}$ can be measured well enough to distinguish between
typical generation non-universal models and generation universal models.
Finally, we found that detector resolution effects 
can degrade $A_{FB}$ measurements quite significantly,  in particular for $t\bar{t}$ final states.
We believe that measurements using 3rd generation fermions look promising 
but clearly, a more careful detector level study is needed to properly take into account
detector resolution and other complications of a real detector. 

\acknowledgments

This research was supported in part by the Natural Sciences and Engineering Research Council 
of Canada.  The authors thank Dag Gillberg, David Morrissey, Dugan O'Neil, Oliver Stelzer-Chilton,
and Isabel Trigger for helpful discussions. 

\begin{widetext}
\begin{table*}[t]
\begin{center}
\caption[ Table of $A_{FB}$ values.] {\setlength{\baselineskip}{0.5cm} 
$A_{FB}$ values for $b$- and $t$-quark final states with corresponding statistical uncertainties, 
assuming $M_{Z^{\prime}}=1.5$~TeV and $L = 100$~fb$^{-1}$.
Cuts include $p_{T}>0.3M_{Z^{\prime}}$~GeV, $|\eta_{t,b}|<2.5$, 
within $|\Delta M_{q\bar{q}} - M_{Z^{\prime}}|<2.5\Gamma_{Z^{\prime}}$. 
The first set of data, labelled as ``ideal detector", do not include detector resolution smearing for 
the final state fermions, while the second set of data does include the effects of detector resolution.}
\vspace{4mm}
\begin{ruledtabular}
\begin{tabular}{ l l  r @{$\;\pm\;$} l l r @{$\;\pm\;$} l l  r @{$\;\pm\;$} l l r @{$\;\pm\;$} l }
\textbf{Model} & $\;\;$ & $A_{FB}^{t}$ & $\delta A_{FB}^{t}$ & $\;\;$ & $A_{FB}^{b}$ & $\delta A_{FB}^{b}$ & $\;\;\;\;\;$ & $A_{FB}^{t}$ & $\delta A_{FB}^{t}$ & $\;\;$ & $A_{FB}^{b}$ & $\delta A_{FB}^{b}$\\ 
\hline
& &\multicolumn{5}{c}{ideal detector} & & \multicolumn{5}{c}{including detector resolution} \\
\hline
$E_{6}\;\chi$	&	&	0.00		&	0.02		&	&	0.060	&	0.011	&	&	0.00		&	0.03		&	&	0.017	&	0.016 \\	
$E_{6}\;\psi$	&	&	0.00		&	0.03		&	&	0.000	&	0.016	&	&	0.00		&	0.05		&	&	0.00		&	0.02\\	
$E_{6}\;\eta$	&	&	0.00		&	0.03		&	&	0.013	&	0.014	&	&	0.00		&	0.04		&	&	0.00		&	0.02\\	
LRM			&	&	0.097	&	0.016	&	&	0.181	&	0.007	&	&	0.04		&	0.02		&	&	0.075	&	0.011\\	
ALRM		&	&	0.146	&	0.019	&	&	0.032	&	0.013	&	&	0.04		&	0.03		&	&	0.006	&	0.016\\	
UUM			&	&	0.144	&	0.007	&	&	0.189	&	0.004	&	&	0.073	&	0.011	&	&	0.094	&	0.006\\	
SSM			&	&	0.049	&	0.014	&	&	0.112	&	0.007	&	&	0.02		&	0.02		&	&	0.044	&	0.010\\	
TC2			&	&	0.069	&	0.017	&	&	0.027	&	0.010	&	&	0.02		&	0.02		&	&	0.01		&	0.01\\	
LH			&	&	0.114	&	0.012	&	&	0.158	&	0.006	&	&	0.047	&	0.018	&	&	0.063	&	0.009\\	
SLH			&	&	0.086	&	0.018	&	&	0.136	&	0.009	&	&	0.03		&	0.03		&	&	0.042	&	0.014\\	
AFSLH		&	&	0.079	&	0.019	&	&	0.124	&	0.010	&	&	0.02		&	0.03		&	&	0.035	&	0.015\\	
3-3-1 (2U 1D)	&	&	0.050	&	0.018	&	&	0.083	&	0.010	&	&	0.02		&	0.03		&	&	0.023	&	0.014\\	
ETC			&	&	0.058	&	0.011	&	&	0.086	&	0.006	&	&	0.022	&	0.015	&	&	0.031	&	0.008\\	
\end{tabular}
\end{ruledtabular}
\label{table:afb}
\end{center}
\end{table*}
\end{widetext}


\end{document}